\documentclass[a4paper]{article}
\usepackage{subfigure}
\usepackage{INTERSPEECH2022}
\usepackage{hyperref}

\title{Improving Visual Speech Enhancement Network by Learning Audio-visual Affinity with Multi-head Attention}
\name{Xinmeng Xu$^{1,2}$, Yang Wang$^2$, Jie Jia$^2$, Binbin Chen$^2$, Dejun Li$^{3,*}$\thanks{$^{*}$Corresponding Author}}
\address{
  $^1$E.E. Engineering, Trinity College Dublin, Ireland\\
  $^2$vivo AI Lab, P.R. China\\
  $^3$School of Foreign Languages, Hubei University of Chinese Medicine, P.R. China}
\email{xux3@tcd.ie,$\{$yang.wang.rj, jie.jia, bb.chen$\}$@vivo.com, dejun$\_$li2022@163.com}

\begin{document}

\maketitle
\begin{abstract}
Audio-visual speech enhancement system is regarded as one of promising solutions for isolating and enhancing speech of desired speaker. Typical methods focus on predicting clean speech spectrum via a naive convolution neural network based encoder-decoder architecture, and these methods a) are not adequate to use data fully, b) are unable to effectively balance audio-visual features. The proposed model alleviates these drawbacks by a) applying a model that fuses audio and visual features layer by layer in encoding phase, and that feeds fused audio-visual features to each corresponding decoder layer, and more importantly, b) introducing a 2-stage multi-head cross attention (MHCA) mechanism to infer audio-visual speech enhancement for balancing the fused audio-visual features and eliminating irrelevant features. This paper proposes attentional audio-visual multi-layer feature fusion model, in which MHCA units are applied to feature mapping at every layer of decoder. The proposed model demonstrates the superior performance of the network against the state-of-the-art models. Speech samples are available at: \scriptsize{\href{https://XinmengXu.github.io/AVSE/AVCRN.html}{\texttt{https://XinmengXu.github.io/AVSE/AVCRN.html}}}
\end{abstract}

\section{Introduction}

Speech processing systems are commonly used in a variety of applications such as automatic speech recognition, speech synthesis, and speaker verification. Numerous speech processing devices (e.g. mobile communication systems and digital hearing aids systems) which are often used in environments with high levels of ambient noise such as public places and cars in our daily life. Generally speaking, the presence of high-level noise interference, severely decreases perceptual quality and intelligibility of speech signal. Therefore, there is the need for the development of speech enhancement algorithms which can automatically filter out noise signal and improve the effectiveness of speech processing systems.

Recently, many methods are proposed to recover the clean speech of target speaker immersed in noisy environment, which can be roughly divided into two categories, i.e., audio-only speech enhancement (AO-SE)\cite{ref1, ref2, ref3} and audio-visual speech enhancement (AV-SE)\cite{ref4, ref5, ref6}. AO-SE methods make assumptions on statistical properties of the involved signals\cite{ref7, ref8}, and aim to estimate target speech signals according to mathematically tractable criteria\cite{ref9, ref10}. Advanced AO-SE methods based on deep learning predict target speech signal directly, but they tend to depart from the knowledge-based modelling. 

Compared with AO-SE methods, AV-SE methods achieve improvement in the performance of intelligibility of speech enhancement due to the visual aspect which can recover some of the suppressed linguistic features, especially, when target speech is corrupted by noise interference\cite{ref11,ref12}. However, AV-SE model should be trained by using data that representatives of settings in which they are deployed. In order to have robust performance in a wide variety of settings, very large AV datasets for training and testing need to be collected. Furthermore, AV-SE is inherently a multi-modal process, and it focuses not only on determining the parameters of a model, but also on the possible fusion architectures\cite{ref13}. Generally, a naive fusion strategy does not allow to control how the information from audio and the visual modalities is fused. As a consequence, the two modalities dominate each other mutually.

\begin{figure*}[t]
  \centering
  \includegraphics[width=0.85\linewidth]{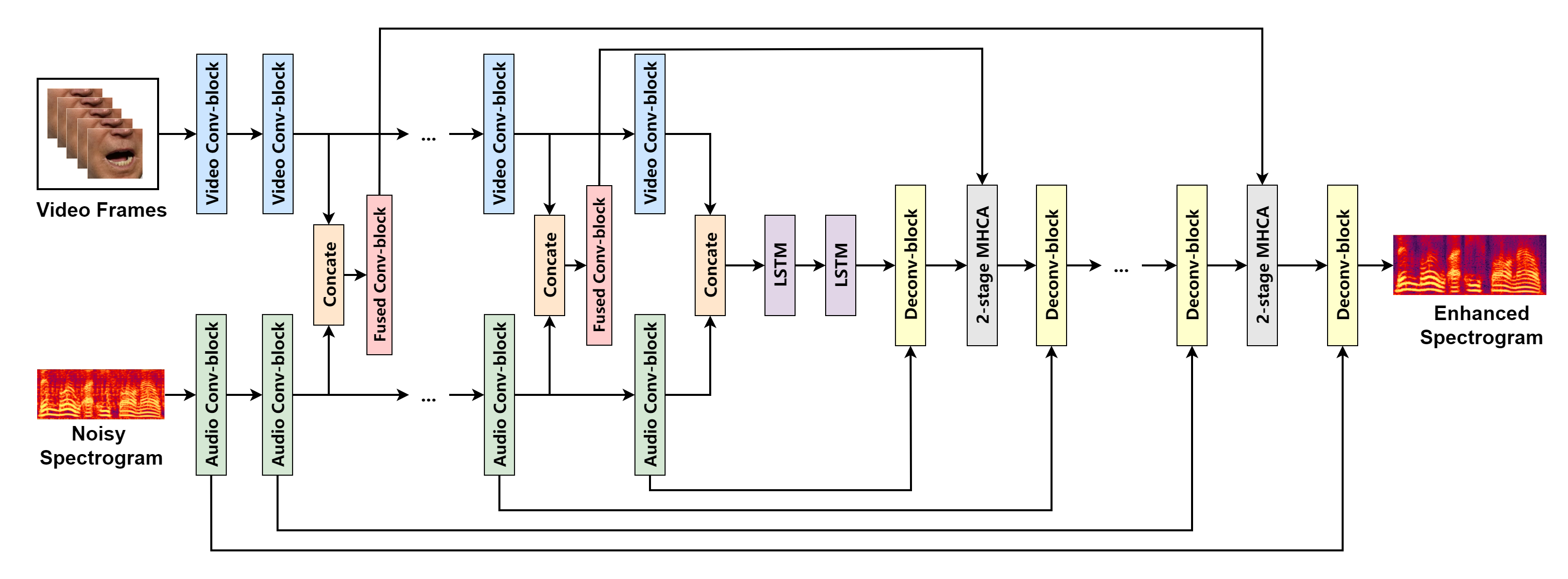}
  \caption{Schematic diagram of the proposed MHCA-AVCRN.}
  \label{fig:0}
\end{figure*}

To adequately use data full and effectively balance audio-visual features, this paper is motivated to design an audio-visual convolutional recurrent network (AVCRN), which can integrate the selected audio and visual cues into a unified network using multi-layer audio-visual fusion strategy while effectively balancing the audio-visual information and filtering out irrelevant features. To this end, this paper proposes an early multi-layer feature fusion based encoder-decoder structure to process the audio and visual input independently and to fuse audio-visual features at each encoder layer for persevering original audio-visual information, and a 2-stage multi-head cross attention (MHCA) mechanism to infer AV-SE for balancing the fused audio-visual features and eliminating irrelevant features. Aggregating this innovation synthetically, we propose MHCA-AVCRN. The contributions of this work are summarized as follows:

\begin{itemize}
    \item A two-stage multi-head cross-attention mechanism is proposed to balance the audio-visual information and filter out irrelevant features to help incorporate audio-visual features and detect interactions between audio mode and visual mode.
    
    \item An efficient early fusion framework for AV-SE is introduced, which extracts audio-visual fused features in different levels and feed them into decoder blocks for promoting the model making better use of data, persevering original audio-visual information, and further boosting the network performance. 
    
    \item Comprehensive studies on publicly available speech data. Results show that MHCA-AVCRN outperforms the evaluated state-of-the-art methods.
\end{itemize}

The reminder of this paper is organised as follows. Section~\ref{sec2} introduces the model architecture. Section~\ref{sec3} illustrates the employed datasets and audio-visual representations. In Section~\ref{sec4} experiment results are presented, and a conclusion is given in Section~\ref{sec5}. 

\section{Model Architecture}\label{sec2}

\subsection{Overview}

The diagram of proposed MHCA-AVCRN is illustrated in Figure~\ref{fig:0}. The model follows an encoder-decoder scheme, and uses a series of downsampling and upsampling blocks to make its prediction. In addition, the proposed MHCA-AVCRN consists of encoder part, fusion part, embedding part, and decoder part, according to Figure~\ref{fig:0}.

The encoder part involves several audio encoder and video encoder blocks. As previous approaches in several convolutional neural networks (CNNs) based audio encoding models \cite{ref14, ref15, ref16}, audio encoder is designed as a CNNs using the Mel-spectrogram as input and each layer of an audio encoder is followed by strided convolutional layer, batch normalization \cite{ref16-1}, and exponential linear unit (ELU) \cite{ref16-2}. The video encoder is used to process the input face embedding through a number of max-pooling convolutional blocks. It is noted that the dimension of visual feature vector after convolution layer has to be the same as the corresponding audio feature vector, since both vectors take at every encoder layer is through a fusion part in encoding stage. Then the encoded features of both streams are concatenated and fed into two LSTM blocks for aggregating temporal contexts. The decoder part is reversed in the audio encoder part by deconvolutions, followed again by batch normalization and ELU. 

Fusion part designates a merged dimension to implement fusion. The audio and video streams of the proposed MHCA-AVCRN take the concatenation operation and are through several strided convolution layer, followed by batch normalization, and ELU. The output of fusion part in each encoder layer and the corresponding decoder layer are fed into the 2-stage MHCA mechanism as is described in Section~\ref{sec:2.2}, for balancing the audio-visual fused features and filtering out irrelevant information from the output of fusion part. 

\subsection{2-stage Multi-head Cross-attention (MHCA)}\label{sec:2.2}

\begin{figure}
  \centering
  \includegraphics[width=0.95\linewidth]{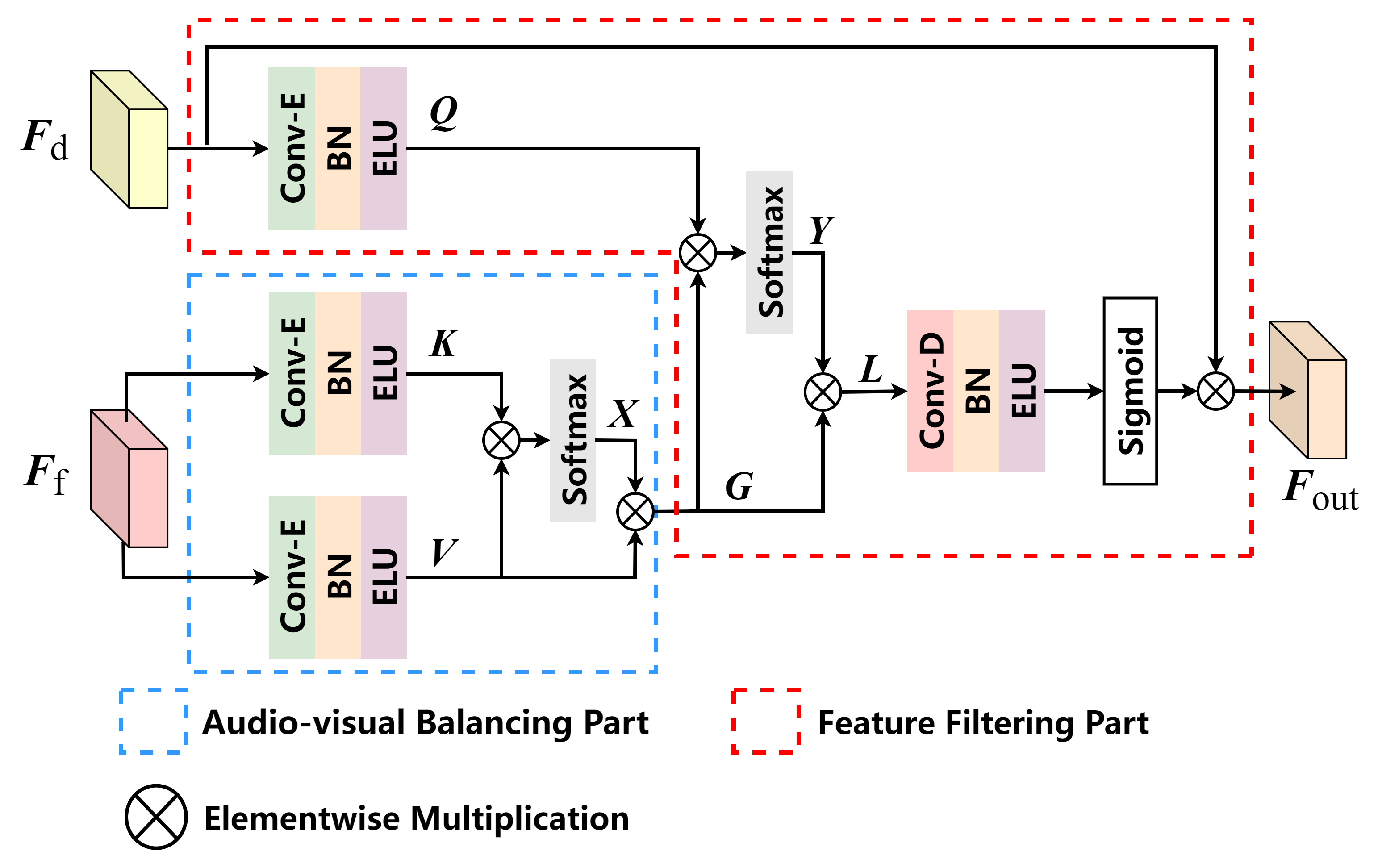}
  \caption{The proposed 2-stage MHCA block, where Conv-E, Conv-D, BN, and LA denote convolution encoder, convolution decoder, and batch normalization, respectively.}
  \label{fig:1}
\end{figure}

The proposed 2-stage MHCA block, as shown in Figure~\ref{fig:1}, is divided into two parts, audio-visual balancing part and the feature filtering part. The 2-stage MHCA block firstly allows the system to attend to the relevant part of audio-visual fused features while mitigating the potential unbalance caused by concatenation-based fusion through audio-visual balancing part. Besides, the 2-stage MHCA block captures the correlation between audio-visual features and audio decoder feature to eliminate the irrelevant information from audio decoder feature through feature filtering part.

In audio-visual balancing part, given an audio-visual fused feature map $F_{\text{f}}$, two 2D convolution layers, each of which is followed by batch normalization and ELU , are adopted to generate new feature map $\textbf{K}$ and $\textbf{V}$ respectively. Concretely, a matrix multiplication between $\textbf{K}$, $\textbf{V}^{\top}$ is operated, to obtain the attention map $\textbf{X}$, a softmax layer is connected as:
\begin{equation}
    x_{ji}=\frac{exp(K_i\times V_j)}{\sum^C_{i=1}exp(K_i\times V_j)},
\end{equation}
in which $x_{ji}$ means the $i^{th}$ channel's influence on the $j^{th}$ channel. Finally, the results of matrix multiplication between $\textbf{X}^{\top}$ and $\textbf{V}$ are weighted by a parameter of scale $\alpha$ to acquire the output of audio-visual balancing part $\textbf{G}$:
\begin{equation}
    G_j=\alpha\sum^C_{i=1}(x_{ji}V_j),
\end{equation}
where $\alpha$ is initialized as zero and can be learned gradually. In feature filtering part, the intermediate feature map $\textbf{G}$ is computed from the output of audio-visual balancing part, $\textbf{G}$ and feature map $\textbf{Q}$ which is computed from audio decoder feature $F_{\text{d}}$ processed by a 2D convolution block.

Next, a matrix multiplication is executed between $\textbf{Q}$ and $\textbf{G}$, and a softmax layer is attached subsequently to calculate the attention maps $\textbf{Y}$:
\begin{equation}
    y_{ji}=\frac{exp(Q_i\times G_j)}{\sum^C_{i=1}exp(Q_i\times G_j)},
\end{equation}
where $y_{ji}$ measures the impact of $i^{th}$ bin to the $j^{th}$ bin. Then, a multiplication of matrix is performed between $\textbf{Y}$ and $\textbf{G}$, and the result is:
\begin{equation}
    L_j=\beta\sum^C_{i=1}(y_{ji}G_j),
\end{equation}
where $\beta$ with a zero initial value can be learned to assign more weight gradually. Finally, the result feature map $\textbf{L}$ is fed into a 2D deconvolution block and a sigmoid function, then is multiplied to $F_{\text{d}}$ to obtain the output of the 2-stage MHCA block, $F_{\text{out}}$.

\section{Datasets and Implementation Details}\label{sec3}
\subsection{Datasets}
The datasets used in the proposed model involve two publicly available audio-visual datasets: GRID\cite{ref22} and TCD-TIMIT\cite{ref23}, which are the two most commonly used databases in the area of audio-visual speech processing. GRID consists of video recordings where 18 male speakers and 16 female speakers pronounce 1000 sentences each. TCD-TIMIT consists of 32 male speakers and 30 female speakers with around 200 videos each.

The proposed model shuffles and splits the dataset to training, validation, and evaluation sets to 24300 (15 males, 12 females, 900 utterance each), 4400 (12 males, 10 females, 200 utterance each), and 1200 utterances (4 males, 4 females, 150 utterance each), respectively. The noise dataset contains 25.3 hours ambient noise categorized into 12 types:  room, car, instrument, engine, train, human chatting, air-brake, water, street, mic-noise, ring-bell, and music. 

Parts of noise signals (23.9 hours) are used in both the training set and the validation set, but the rest are mixed to create the evaluation set. The speech-noise mixtures in training and validation are generated by randomly selecting utterances from speech dataset and noise dataset then mising them up at random SNR between -10dB and 10dB. The evaluation set is generated SNR at 0dB and -5dB.

\subsection{Audio representation}
The audio representation is the transformed magnitude spectrograms in the log Mel-domain. The input audio signals are raw waveforms, and firstly are transformed to spectrograms by using Short Time Fourier Transform (STFT) with Hanning window function, and 16 kHz sampling rate.  Each frame contains a window of 40 milliseconds, which equals 640 samples per frame and corresponds to the duration of a single video frame, and the frame shift is 160 samples (10 milliseconds). 

The transformed spectrograms are then converted to log Mel-scale spectrograms via Mel-scale filter banks. The resulting spectrogram have 80 Mel frequency bands from 0 to 8 kHz. The whole spectrograms are sliced into pieces of duration of 200 milliseconds corresponding to the length of 5 video frames, resulting in spectrograms of size 80$\times$20, representing 20 temporal samples, and 80 frequency bins in each sample.

\subsection{Visual representation}
Visual representation is extracted from the input videos, and is re-sampled to 25 frames per second. Each video is divided into non-overlapping segments of 5 frames. During the processing stage, each frame that has been cropped a mouthcentered window of size 128 × 128 by using the 20 mouth landmarks from 68 facial landmarks suggested by Kazemi et al.\cite{ref24}. Then the video segment processed as input is the size of 128$\times$128$\times$5, and then zoomed to 80$\times$80$\times$5.

\subsection{Network Training}

The models are trained with the Adam optimizer \cite{ref25}. We set the learning rate to 0.0002. The mean squared error (MSE) serves as the objective function. We train the models with a minibatch size of 16 on the utterance-level. Within a minibatch, all training samples are padded with zeros to have the same number of time steps as the longest sample does. The best models are selected by cross validation.

\subsection{Evaluation Metrics}
The following two metrics are used to evaluate MHCA-AVCRN and state-of-the-art competitors. All metrics are better if higher.
\begin{itemize}
    \item PESQ: Perceptual evaluation of speech quality (from $-0.5 $ to $4.5$) \cite{ref25-1}.
    \item STOI: Short-time objective intelligibility measure (from $0$ to $100(\%)$) \cite{ref25-2}.
\end{itemize}

\section{Experiment Results}\label{sec4}

\renewcommand{\arraystretch}{0.85}
\begin{table*}[]
\centering
\caption{Evaluation results of proposed model compared with baseline models. Boldface characters indicates the best result of each column.}
\resizebox{\textwidth}{!}{
\begin{tabular}{|c|cccc|cccc|}
\hline
Evaluation metrics  & \multicolumn{4}{c|}{STOI(\%)}                                                                                                    & \multicolumn{4}{c|}{PESQ}                                                                                                    \\ \hline
Test SNR            & \multicolumn{2}{c|}{-5 dB}                                                & \multicolumn{2}{c|}{0 dB}                            & \multicolumn{2}{c|}{-5 dB}                                              & \multicolumn{2}{c|}{0 dB}                          \\ \hline
Interference        & \multicolumn{1}{c|}{Speech}         & \multicolumn{1}{c|}{Natural}        & \multicolumn{1}{c|}{Speech}         & Natural        & \multicolumn{1}{c|}{Speech}        & \multicolumn{1}{c|}{Natural}       & \multicolumn{1}{c|}{Speech}        & Natural       \\ \hline
Unprocessed         & \multicolumn{1}{c|}{57.82}          & \multicolumn{1}{c|}{51.44}          & \multicolumn{1}{c|}{64.74}          & 62.59          & \multicolumn{1}{c|}{1.59}          & \multicolumn{1}{c|}{1.03}          & \multicolumn{1}{c|}{1.66}          & 1.24          \\ \hline
Audio-only CRN \cite{ref26}     & \multicolumn{1}{c|}{72.26}          & \multicolumn{1}{c|}{78.67}          & \multicolumn{1}{c|}{80.82}          & 83.35          & \multicolumn{1}{c|}{2.01}          & \multicolumn{1}{c|}{2.19}          & \multicolumn{1}{c|}{2.47}          & 2.58          \\
VSE \cite{ref27}                & \multicolumn{1}{c|}{79.94}          & \multicolumn{1}{c|}{83.33}          & \multicolumn{1}{c|}{84.61}          & 87.90          & \multicolumn{1}{c|}{2.36}          & \multicolumn{1}{c|}{2.41}          & \multicolumn{1}{c|}{2.77}          & 2.94          \\
AV(SE)$^2$ \cite{ref28}             & \multicolumn{1}{c|}{81.06}          & \multicolumn{1}{c|}{85.41}          & \multicolumn{1}{c|}{86.17}          & 89.44          & \multicolumn{1}{c|}{2.42}          & \multicolumn{1}{c|}{2.49}          & \multicolumn{1}{c|}{2.81}          & 2.98          \\ \hline
MHCA-AVCRN (Proposed) & \multicolumn{1}{c|}{\textbf{82.22}} & \multicolumn{1}{c|}{\textbf{86.89}} & \multicolumn{1}{c|}{\textbf{88.37}} & \textbf{90.10} & \multicolumn{1}{c|}{\textbf{2.64}} & \multicolumn{1}{c|}{\textbf{2.72}} & \multicolumn{1}{c|}{\textbf{2.95}} & \textbf{3.08} \\
\quad\quad-w/o 2-stage MHCA   & \multicolumn{1}{c|}{80.47}          & \multicolumn{1}{c|}{82.71}          & \multicolumn{1}{c|}{85.02}          & 88.94          & \multicolumn{1}{c|}{2.39}          & \multicolumn{1}{c|}{2.47}          & \multicolumn{1}{c|}{2.83}          & 2.96          \\ \hline
\end{tabular}}
\label{tab:1}
\end{table*}

\renewcommand{\arraystretch}{0.85}
\begin{table*}[]
\centering
\caption{Evaluation results between ``incomplete MHCA-AVCRN'' and the ``complete MHCA-AVCRN''. \textbf{BOLD} indicates the best result of each column. }
\resizebox{\textwidth}{!}{
\begin{tabular}{|c|cccc|cccc|}
\hline
Evaluation metrics                         & \multicolumn{4}{c|}{STOI(\%)}                                                                                                    & \multicolumn{4}{c|}{PESQ}                                                                                                    \\ \hline
Test SNR                                   & \multicolumn{2}{c|}{-5 dB}                                                & \multicolumn{2}{c|}{0 dB}                            & \multicolumn{2}{c|}{-5 dB}                                              & \multicolumn{2}{c|}{0 dB}                          \\ \hline
Interference                               & \multicolumn{1}{c|}{Speech}         & \multicolumn{1}{c|}{Natural}        & \multicolumn{1}{c|}{Speech}         & Natural        & \multicolumn{1}{c|}{Speech}        & \multicolumn{1}{c|}{Natural}       & \multicolumn{1}{c|}{Speech}        & Natural       \\ \hline
Unprocessed                                & \multicolumn{1}{c|}{57.82}          & \multicolumn{1}{c|}{51.44}          & \multicolumn{1}{c|}{64.74}          & 62.59          & \multicolumn{1}{c|}{1.59}          & \multicolumn{1}{c|}{1.03}          & \multicolumn{1}{c|}{1.66}          & 1.24          \\ \hline
MHCA-AVCRN w/o 2-stage MHCA                & \multicolumn{1}{c|}{80.47}          & \multicolumn{1}{c|}{82.71}          & \multicolumn{1}{c|}{85.02}          & 88.94          & \multicolumn{1}{c|}{2.39}          & \multicolumn{1}{c|}{2.47}          & \multicolumn{1}{c|}{2.83}          & 2.94          \\
MHCA-AVCRN w/o audio-visual balancing part & \multicolumn{1}{c|}{80.98}          & \multicolumn{1}{c|}{84.84}          & \multicolumn{1}{c|}{86.23}          & 89.02          & \multicolumn{1}{c|}{2.48}          & \multicolumn{1}{c|}{2.58}          & \multicolumn{1}{c|}{2.87}          & 2.96          \\
MHCA-AVCRN w/o feature filtering part      & \multicolumn{1}{c|}{81.56}          & \multicolumn{1}{c|}{85.22}          & \multicolumn{1}{c|}{87.07}          & 89.78          & \multicolumn{1}{c|}{2.52}          & \multicolumn{1}{c|}{2.60}          & \multicolumn{1}{c|}{2.91}          & 3.01          \\ \hline
MHCA-AVCRN                                 & \multicolumn{1}{c|}{\textbf{82.22}} & \multicolumn{1}{c|}{\textbf{86.89}} & \multicolumn{1}{c|}{\textbf{88.37}} & \textbf{90.10} & \multicolumn{1}{c|}{\textbf{2.64}} & \multicolumn{1}{c|}{\textbf{2.72}} & \multicolumn{1}{c|}{\textbf{2.95}} & \textbf{3.08} \\ \hline
\end{tabular}}
\label{tab:2}
\end{table*}

\subsection{Overall Performance}
Table\ref{tab:1} presents comprehensive evaluations for three baseline models on untrained noises and untrained speakers. The best results in each case are highlighted in boldface. In addition, the three selected baseline models are: (1) \textbf{Audio-only CRN} \cite{ref26}, an audio-only speech enhancement model based on CRN, (2) \textbf{VSE} \cite{ref27}, an audio-visual neural network for visual speech enhancement, and (3) \textbf{AV(SE)$^2$}, an audio-visual speech enhancement model with several cross-modal squeeze-excitation blocks.

Table~\ref{tab:1} demonstrates the improvement in the performance of networks, as new component to the speech enhancement architecture, such as visual modality, audio-visual early feature fusion strategy, and finally the proposed 2-stage MHCA. There are several observations that (1) the VSE outperforms the audio-only CRN, (2) the proposed MHCA-AVCRN without 2-stage MHCA performs better than VSE, and (3) the performance of proposed MHCA-AVCRN better than the AV(SE)$^2$ which is also an attention based AV-SE model but uses late feature fusion strategy. Hence the performance improvement from audio-only CRN to proposed MHCA-AVCRN is primarily for three reasons: (I) the addition of the visual modality, (II) the use of fusion technique of audio-visual early feature fusion strategy, rather than concatenating audio and visual modalities only once in the whole network, and (III) the use of 2-stage MHCA that improves the performance of the proposed model further. 

\begin{figure}[]
    \centering
    \subfigure[Noisy input]{
    \begin{minipage}[b]{0.22\textwidth}
    \includegraphics[width=1\textwidth]{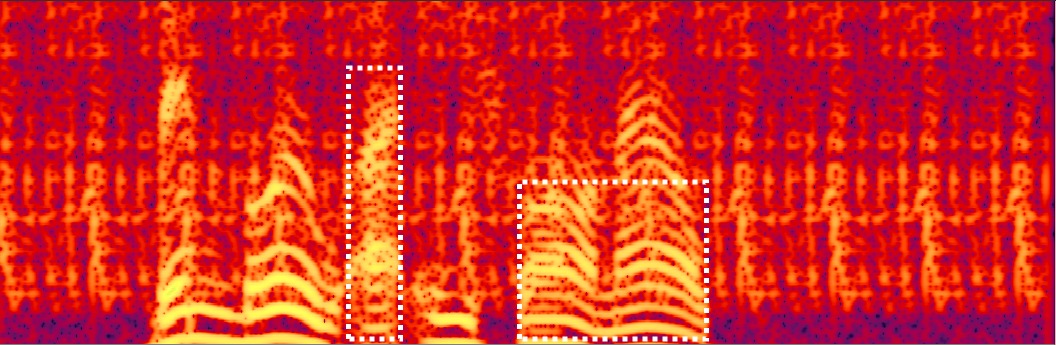}
    \end{minipage}
    }
    \subfigure[VSE enhanced]{
    \begin{minipage}[b]{0.22\textwidth}
    \includegraphics[width=1\textwidth]{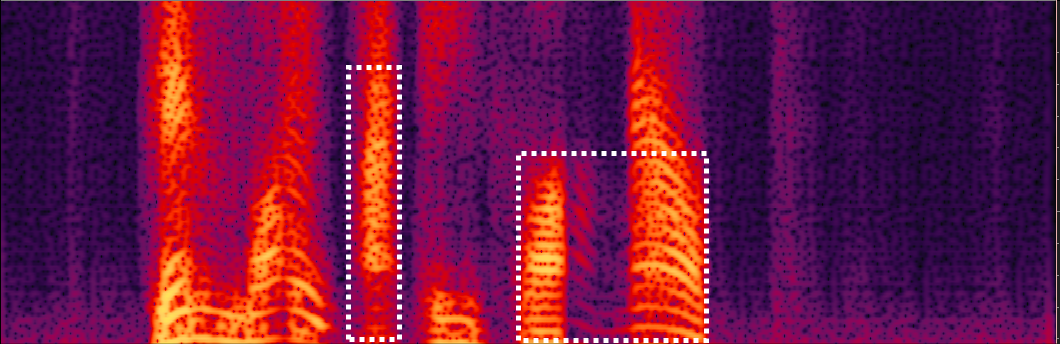}
    \end{minipage}
    }\\
     \subfigure[AV(SE)$^2$ enhanced]{
    \begin{minipage}[b]{0.22\textwidth}
    \includegraphics[width=1\textwidth]{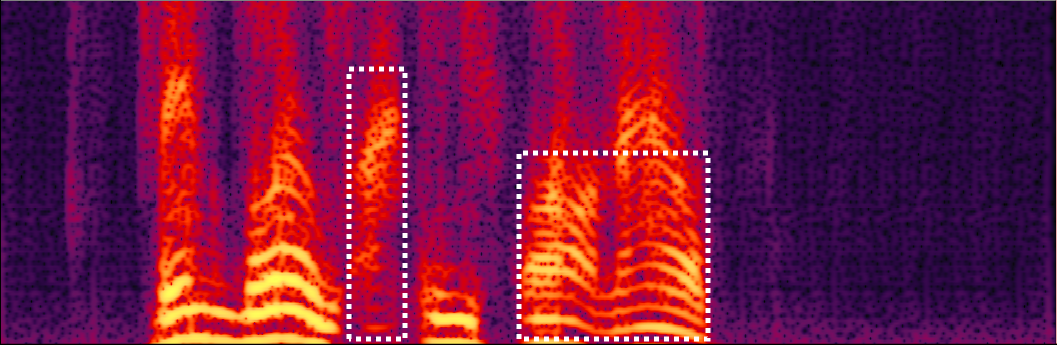}
    \end{minipage}
    }
     \subfigure[MHCA-AVCRN enhanced]{
    \begin{minipage}[b]{0.22\textwidth}
    \includegraphics[width=1\textwidth]{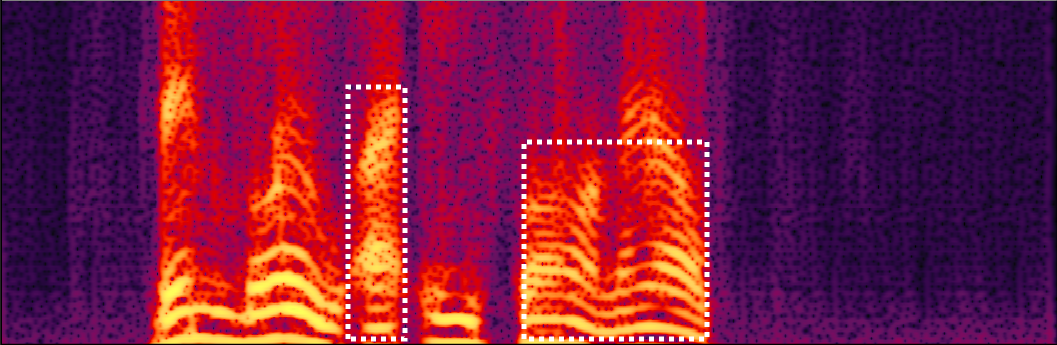}
    \end{minipage}
    }
    \caption{Example of input and enhanced spectra from an example speech utterance.} \label{fig3}
\end{figure}

Figure~\ref{fig3}  shows a visualization of VSE enhancement, AV(SE)2 enhancement, and the proposed MHCA-AVCRN enhancement. For comparison details of the spectra are framed by dotted boxes.

\subsection{Impact of 2-stage MHCA}

In this study, we evaluate variants of proposed MHCA-AVCRN when removing audio-visual balancing part, feature filtering part, and 2-stage MHCA respectively, and compare the performance between these three ``incomplete MHCA-AVCRN'' and the ``complete MHCA-AVCRN''.

From Table~\ref{tab:2}, we can conclude that the existence of the audio-visual balancing part and the feature filtering part do promote the network performance. The 2-stage MHCAs improves $4.18\%$ and $0.25$ on STOI and PESQ respectively under the condition of natural noise at -5 dB. Furthermore, a single audio-visual balancing part ($2.51\%$ on STOI and $0.13$ on PESQ under the condition of natural noise at -5 dB) performs better than a single feature filtering part ($2.13\%$ on STOI and $0.11$ on PESQ under the condition of natural noise at -5 dB).

\renewcommand{\arraystretch}{0.9}
\begin{table}[]
\caption{Performance comparison of MHCA-AVCRN with state-of-the-art result under natural noise on GRID.}
\centering
\begin{tabular}{|c|c|c|c|c|}
\hline
Test SNR                          & \multicolumn{2}{c|}{-5 dB} & \multicolumn{2}{c|}{0 dB} \\ \hline
Evaluation Metrics                   & \multicolumn{4}{c|}{$\Delta$PESQ}                              \\ \hline
Deep-learning-based AV-SE & \multicolumn{2}{c|}{1.13}  & \multicolumn{2}{c|}{0.74} \\ 
OVA Approach              & \multicolumn{2}{c|}{0.21}  & \multicolumn{2}{c|}{0.06} \\ 
L2L Model                 & \multicolumn{2}{c|}{0.26}  & \multicolumn{2}{c|}{0.19} \\ \hline
\end{tabular}
\label{tab:3}
\end{table}

\subsection{Comparison with Reference Models}

Table~\ref{tab:3} demonstrated that our proposed approach produces state-of-the-art results in terms of speech quality metrics as discussed above by comparing against the following three recently proposed methods that use deep neural networks to perform AV-SE on GRID dataset:
\begin{itemize}
    \item
    \textbf{Deep-learning-based AV-SE}\cite{ref29}: Deep-learning-based audio-visual speech enhancement in presence of Lombard effect
    \item
    \textbf{OVA approach}\cite{ref30}: An LSTM based AV-SE with mask estimation
    \item
    \textbf{L2L model}\cite{ref31}: A speaker independent audio-visual model for speech separation
\end{itemize}

The results where $\Delta$PESQ denotes PESQ improvement between the selected models result in reference papers and the MHCA-AVCRN result in Table~\ref{tab:1}. Results for the competing methods are taken from the corresponding papers. Although the comparison results are for reference only, the proposed model demonstrates a robust performance in comparison with state-of-the-art results on the GRID AV-SE tasks.

\section{Conclusions}\label{sec5}

This paper proposes an MHCA-AVCRN model for audio-visual speech enhancement. The early multi-layer feature fusion strategy repeat downsampling and convolution of feature maps to combine both high-level and low-level features at different layer steps. In addition, a 2-stage MHCA mechanism which contains the audio-visual balancing part and feature filtering part to infer AV-SE is introduced for balancing the fused audio-visual features and eliminating irrelevant features. Results provide an illustration that the proposed model has better performance than some published state-of-the-art models on the GRID dataset.

\bibliographystyle{IEEEtran}

\bibliography{mybib}


\end{document}